\documentclass[12pt]{iopart}

\begin{document}

\title[Exact dynamics of XX central spin models]{Exact dynamics of XX central spin models}

\author{M. A. Jivulescu $^{ 1,2}$, E. Ferraro $^1$, A. Napoli $^1$, A. Messina $^1$}

\address{$^1$ CNISM and Dipartimento di Scienze
Fisiche ed Astronomiche,  Universit\`{a} di Palermo, Italy\\
$^2$ Department of Mathematics, "Politehnica" University of
Timi\c{s}oara, Romania}
\ead{maria.jivulescu@mat.upt.ro;messina@fisica.unipa.it}
\begin{abstract}

The dynamical behavior of a star network of spins, wherein each of
$N$ decoupled  spins interact  with a central spin through non
uniform Heisenberg XX interaction is exactly studied. The
time-dependent Schr\"{o}dinger equation of the spin system model
is solved starting from an arbitrary initial state. The resulting
solution is analyzed and briefly discussed.

\end{abstract}

%\vspace{2pc} \noindent{\it Keywords}: central spin model,
\maketitle
\section{Introduction}
 In the last decades investigations on the properties of
coupled-spin systems has gained an increasing interest in  the
quantum community\cite{semicond}. Specially, the time evolution of
a spin star system, that is a single spin coupled to a surrounding
environment composed by a
finite number of spins noninteracting \cite{Woods}%\cite{Rajagopal}, \cite{Breuer},
-\cite{Ferraro} or interacting\cite{Deng}-\cite{Petruccione} among
them, has been studied in detail. Central spin models provide an
appropriate description of  quantum information processes such as,
for instance, quantum state transfer\cite{Bose} and quantum
cloning\cite{Chiara}.

In this paper, we study in detail the dynamics of a XX central
spin model that is composed by a localized spin 1/2, hereafter
called central spin, coupled to an environment of $N$ not
interacting ½ spins. The Hamiltonian  describing our system is
\begin{equation}\label{ham}
H=\omega\sum_{j=1}^N\sigma_z^j+\omega_0\sigma_z^A+\sum_{j=1}^N\alpha_j(\sigma_+^A\sigma_-^j+\sigma_-^A\sigma_+^j).
\end{equation}
The Pauli operators  $\sigma_{\pm}^A$ refer to the central spin
whereas the others, labelled by the index $j$, refer to the $N$
environmental  spins. The central spin Hamiltonian model
(\ref{ham}) can be successfully exploited to effectively describe
many physical systems like quantum dots\cite{Imamoglu},
two-dimensional electron gases\cite{pr} and optical
lattices\cite{so}. The Hamiltonian model given by eq. (\ref{ham})
is a realization of the so-called Gaudin model whose
diagonalization has been derived in the framework of the Bethe
ansatz (BA) \cite{Gaudin}. Such an approach provides however a
rather formal solution whose implications for the dynamics of the
spin system have not  yet fully explored. In this paper we solve
exactly the Schr\"{o}dinger equation of motion of the total system
starting from an arbitrary completely factorized initial
condition. Our successfully treatment  is strictly related to the
circumstance that the component along the $z$ axes of the total
angular momentum operator
$S_z=\frac{\sigma_z^A}{2}+\frac{1}{2}\sum_{j=1}^N\sigma_z^j\equiv\frac{\sigma_z^A}{2}+J_z$
is a constant of motion. Our main result is the derivation of a
closed formula for the time dependence of the probability
amplitude of finding our spin system in any given state.
\section{Exact dynamics of XX-central spin model} The goal of this
Section is to derive the exact dynamics of our spin system
starting from a  completely factorized initial condition wherein
the central spin, as well as $p$ $(p=0,\ldots, N-1)$ of the $N$
surrounding spins are in their respective up state, whereas the
others ones are prepared in their down state. The case
corresponding to $p=N$ is  trivial since the corresponding
factorized state  is an eigenstate of the Hamiltonian (\ref{ham}).
\subsection{ p=0} The initial condition taken into consideration in
this subsection is the following one:
\begin{equation}\label{p=0}
|\psi(0)\rangle=|\!\!\uparrow_A\rangle|
\downarrow\ldots\downarrow\rangle,
\end{equation}
where only the central spin is in the up state. It is easy to
convince oneself \cite{Ferraro} that, thanks to the conservation
of $S_z$, the state (\ref{p=0}) evolves into the state
 representable by the following  normalized
superposition
$|\psi(t)\rangle=a(t)|\uparrow_A\rangle|\!\!\downarrow
\ldots\downarrow\rangle+\sum\limits_{j=1}^Nb_j(t)|\!\!\downarrow_A\rangle|\downarrow\
\ldots\uparrow_j\downarrow\rangle, $ where
\begin{equation}\label{a4}
a(t)=\cos\left(\sqrt{\sum_{j=1}^N\alpha_j^2+\Delta^2}\;t\right)-i\frac{\Delta}{\sqrt{\sum_{j=1}^N\alpha_j^2+\Delta^2}}\sin\left(\sqrt{\sum_{j=1}^N\alpha_j^2+\Delta^2}\;t\right)
\end{equation}
\begin{equation}\label{b4}
b_j(t)=-i\frac{\alpha_j}{\sqrt{\sum_{j=1}^N\alpha_j^2+\Delta^2}}\sin\left(\sqrt{\sum_{j=1}^N\alpha_j^2+\Delta^2}\;t\right).
\end{equation}
with $\Delta=\omega-\omega_0$.  We underline that starting from
such an initial condition, the time evolution is characterized by
only one effective frequency, namely
$\alpha_{eff}=\sqrt{\sum_{j=1}^N\alpha_j^2+\Delta^2}$. Thus the
spin system fully restores its initial condition with a period
$T=2\pi/\alpha_{eff}$ and behaves as if its dynamics were governed
by an effective Hamiltonian model like that one given by eq.
(\ref{ham}), where $\alpha_j$ is substituted by $\alpha_{eff}$,
independent of $j$.
\subsection{p=1,2\ldots, N-1}
Our aim is now to treat the more complicated dynamics of the XX
central spin system starting from an arbitrary initial condition
of the form
\begin{equation}\label{condizione}
|\psi(0)\rangle=|\!\!\uparrow_A\rangle|\!\!\downarrow\ldots\uparrow_{k_1}
\ldots\uparrow_{k_p}\ldots\downarrow\rangle,
\end{equation}
where $p$ $(p\neq0$) of the $N$ uncoupled spins, labelled by
$k_1,\ldots,k_p$, are in their up state $|\uparrow\rangle$, while
the remaining $N-p$ spins are in their down state
$|\!\!\downarrow\rangle$. Since $[S_z,H]=0$, we claim that at any
time instant $t$ the system evolves
into a normalized superposition of $\left(\begin{array}{c}N\\
p\end{array} \right)\equiv C_N^p$ states wherein the central spin,
as well as $p$ among
the $N$, are up and $\left(\begin{array}{c} N\\
p+1 \end{array} \right)\equiv C_N^{p+1}$ states wherein the
central spin is down and $p+1$ spins among the $N$ are up. Thus,
starting from the initial condition (\ref{condizione}), the vector
state of the system evolves within a finite dimensional subspace
whose dimension is $C_N^p+C_N^{p+1}=C_{N+1}^{p+1}$. Starting from
this property we proceed to write down effectively the evolved
state of the system. To represent it, we exploit the set of
$p$-tuples, that is the set of all the subsets of $p$ elements
from the first $N$ natural numbers,
$\mathcal{S}_p=\{(i_1,i_2,\ldots,i_p), 1\leq i_1<\ldots<i_p\leq N
\}$. It is well-known that the number of all $p$-tuples from $N$
numbers is exactly $C_N^p$. Therefore, establishing a bijection
between the set $\mathcal{S}_p$ and the set of the states
$\{|\uparrow_A\rangle|\!\!\downarrow\uparrow_{i_1}
\ldots\uparrow_{i_p}\ldots\downarrow\rangle\}$, as well as between
the set $\mathcal{S}_{p+1}$ and the set
$\{|\downarrow_A\rangle|\!\!\downarrow\ldots\uparrow_{j_1}
\ldots\downarrow\uparrow_{j_{p+1}}\rangle\}$,  it is possible to
represent the state of the system at the generic time instant $t$
as follows:
\begin{eqnarray}\label{statot}\hspace{-2cm}
|\psi(t)\rangle=&\sum\limits_{\mathcal{S}_p}a_{(i_1,i_2,\ldots,i_p)}(t)|\uparrow_A\rangle|\!\!\downarrow\uparrow_{i_1}
\ldots\uparrow_{i_p}\rangle
&+\sum\limits_{\mathcal{S}_{p+1}}b_{(j_1,j_2,\ldots,j_{p+1})}(t)|\!\!\downarrow_A\rangle|\!\!\downarrow\uparrow_{j_1}\ldots
\uparrow_{j_{p+1}}\rangle,
\end{eqnarray}
where the $p$-tuple $(i_1,i_2,\ldots,i_p)\in \mathcal{S}_p$
identifies the probability amplitude $a_{(i_1,i_2,\ldots,i_p)}(t)$
of finding central spin $A$ and exactly the spins $i_1,i_2,\ldots,
i_p$  among the $N$ around spins in their respective up state.
Analogously, the $(p+1)$-tuple $(j_1,j_2,\ldots, j_{p+1})\in
\mathcal{S}_{p+1}$ provides the probability amplitude
$b_{(j_1,j_2,\ldots,j_{p+1})}(t)$  of finding out the spin system
in the particular state with the central spin down and exactly the
environmental spins $j_1,j_2,\ldots, j_{p+1}$ up. In order to get
explicit equations for $\{a_{(i_1,i_2,\ldots,i_p)}(t)\}$ and
$\{b_{(j_1,j_2,\ldots,j_{p+1})}(t)\}$ we start from the
time-dependent Schr\"{o}dinger equation, introducing  an
appropriate mathematical notation useful to represent the
transformations undergone by the states appearing in the
expression (\ref{statot}) by the application of the Hamiltonian
(\ref{ham}). For this reason, we define two families of mappings
$\{O_r\}_{r=1}^N$ and $\{\delta_r\}_{r=1}^N$. For any fixed $r$,
the mapping $O_r$ transforms a $p$-tuple into a $(p+1)$-tuple
accordingly to the rule
\begin{eqnarray} O_r:\mathcal{S}_{p}^{-\{ r\}}\rightarrow \mathcal{S}_{p+1}^{ \cup\{ r\}},\quad O_r(i_1,i_2,\ldots, i_p)=(i_1,i_2,\ldots,i_p)\cup\{r\}.\end{eqnarray}
where $\mathcal{S}_{p}^{-\{ r\}}(\mathcal{S}_{p}^{\cup\{ r\}})$
represents the set of all subsets of $p$ elements, diverse by $r$
(including $r$), from the the first $N$ naturals numbers.
 The mapping $O_r$ adds the natural number $r$ to the $p$-tuple
$(i_1,\ldots,i_p)$, arranging them in increasing order. We point
out that this correspondence is well defined if and only if  $r$
does not belong to the set $\{i_1,i_2,\ldots,i_p\}$. The family of
mappings $\{\delta_r\}_{r=1}^n$ on the contrary transforms, for
any fixed $r$, a $(p+1)$-tuple in a $p$-tuple in accordance with
\begin{eqnarray} \delta_r: \mathcal{S}_{{p+1}}^{\cup\{ r\}}\rightarrow \mathcal{S}_{p}^{-\{ r\}},\quad \delta_r(j_1,j_2,\ldots, j_{p+1})=(j_1,j_2,\ldots,j_{p+1})-\{r\}.\end{eqnarray}
It acts on the family of $p+1$ elements, recovering a $p$-tuple
from $\{j_1,j_2,\ldots,j_{p+1}\}$ by eliminating the element $r$.
Obviously the above correspondence is well defined if and only if
$r$ belongs to the set $\{j_1,j_2,\ldots,j_{p+1}\}$. Inserting
eq.(\ref{statot}) into the time-dependent Schr\"{o}dinger
equation,  we obtain the following system of coupled equations for
the probability amplitudes $a_{(i_1,i_2,\ldots,i_p)}(t)$ and
$b_{(j_1,j_2,\ldots,j_{p+1})}(t)$
\begin{equation}\hspace{-1cm}\label{a1}
i\,\dot{a}_{(i_1,i_2,\ldots,i_p)}(t)=\Delta\,a_{(i_1,i_2,\ldots,i_p)}(t)+\sum_{r=1
 (r\in\!\!\!/\{ i_1, i_2,\dots,
 i_p\})}^N\alpha_rb_{O_r(i_1,\ldots,i_p)}(t)\\
\end{equation}
\begin{equation}\hspace{-1cm}\label{b1}
i\,\dot{b}_{(j_1,j_2,\ldots,j_{p+1})}(t)=-\Delta\,
b_{(j_1,j_2,\ldots,j_{p+1})}(t)+\sum_{r=1 (r\in\{ j_1, j_2,\dots,
j_{p+1}\})}^N\alpha_ra_{\delta r(j_1,\ldots,j_{p+1})}(t).
\end{equation}
Solving the above system requires  the diagonalization of the
companion matrix of the system which is of order $C_{N+1}^{p+1}$.
On the other hand, we notice that the mean value of $<S_z^A>$ may
be expressed as
$<S_z^A>=\sum\limits_{\mathcal{S}_p}|a_{(i_1,\dots,i_p)}|^2-1/2=1/2-\sum\limits_{\mathcal{S}_{p+1}}|b_{(i_1,\dots,i_{p+1})}|^2$,
so that to decouple the system of eqs. (\ref {a1})-(\ref{b1}) is
of physical and mathematical interest. Thus, to proceed further we
follow a standard procedure by which we succeed in converting the
above system into two decoupled systems for each unknown set
$\{a_{(i_1,i_2,\ldots,i_p)}(t)\}$ and
$\{b_{(j_1,j_2,\ldots,j_{p+1})}(t)\}$. In this way we get two
systems of coupled equations for the amplitudes
$a_{(i_1,i_2,\ldots,i_p)}(t)$ and
$b_{(j_1,j_2,\ldots,j_{p+1})}(t)$ respectively. We have indeed
\begin{equation}\hspace{-2cm}\label{a2}
\ddot{a}_{(i_1,i_2,\ldots,i_p)}(t)=-\left[\left(\Delta^2+\sum_{j=p+1}^N\alpha_{i_j}^2\right)a_{(i_1,i_2,\ldots,i_p)}(t)+\sum_{r,s=1\\r\neq
s}^N\alpha_r\alpha_s\,a_{\delta_
s(O_r(i_1,\ldots,i_p))}(t)\right],
\end{equation}
\begin{equation}\hspace{-2.5cm}\label{b2}
\ddot{b}_{(j_1,j_2,\ldots,j_{p+1})}(t)=-\left[\left(\Delta^2+\sum_{i=1}^{p+1}\alpha_{j_i}^2\right)b_{(j_1,j_2,\ldots,j_{p+1})}(t)+\sum_{r,s=1\\r\neq
s}^N\alpha_r\alpha_s\,b_{O_s(\delta_
r(j_1,\ldots,j_{p+1}))}(t)\right]
\end{equation}
Eq.(\ref{a2}) (Eq.(\ref{b2})) defines a linear system of $C_N^p$
($C_N^{p+1}$) coupled second order differential equations in the
variables $a_{(i_1,i_2,\ldots,i_p)}(t)$
($b_{(j_1,j_2,\ldots,j_{p+1})}(t)$).
 The $C_N^p (C_N^{p+1}) $
amplitudes $a_{(i_1,i_2,\ldots,i_p)}(t)$,
($b_{(i_1,i_2,\ldots,i_{p+1})}(t)$) may be ordered in accordance
with lexicographical prescription, that is
$a_{(i'_1,\ldots,i'_p)}$ ($ b_{(i'_1,\ldots,i'_{p+1})}$) follows
$a_{(i_1,\ldots,i_p)}$ ($ b_{(i_1,\ldots,i_{p+1})}$) if $i_1=i'_1,
i_2=i'_2,\dots, i_{m-1}=i'_{m-1},i_m<i'_m$ with $m=1,2,\dots, p$
$(m=1,2,\dots, p+1)$. Therefore, eqs.(\ref{a2}), ((\ref{b2}))
admits the matrix representation
\begin{equation}\label{A}
\ddot{\mathbf{x}}(t)=-\mathcal{X}\,\mathbf{x}(t)
\end{equation}
where $\mathbf{x}(t)=\mathbf{a}(t) (\mathbf{b}(t)) $ is the
lexicographically ordered vector of the probability amplitudes
$a_{(i_1,i_2,\ldots,i_p)}(t)$, $(b_{(j_1,j_2,\ldots,j_{p+1})}(t))$
and $\mathcal{X}=\mathcal{A},(\mathcal{B})$ is the corresponding
companion matrix. In accord with eq. (\ref{a2}), the matrix
elements of $\mathcal{A}=(\mathcal{A}_{mm'})_{1\leq m,m'\leq
C_N^p}$ are given by
\begin{equation}\hspace{-1cm}\label{Amatrix}
\mathcal{A}_{mm'}= \left\{\begin{array}{c}
\Delta^2+\sum\limits_{j=p+1}^N\alpha_{i_j}^2,\quad \quad \quad if
 \quad card (\{m-m'\})=0 (\Leftrightarrow m=m')\\
\alpha_{m-m'}\alpha_{m'-m},\qquad if \quad card (\{m-m'\})=1\\
0, \quad otherwise.
\end{array}\right.,
\end{equation}
where $m=(i_1,i_2,\ldots,i_p)$ and $card (\{m-m'\})$ is the total
number of the elements in the difference set $\{m-m'\}$. In a
similar manner, denoting now the $(p+1)$-tuples by
$q=(j_1,\ldots,j_{p+1})$, the matrix
$\mathcal{B}=(\mathcal{B}_{qq'})_{1\leq q,q'\leq C_N^{p+1}}$ is
defined by
\begin{equation}
\mathcal{B}_{qq'}= \left\{\begin{array}{c}
\Delta^2+\sum\limits_{i=1}^{p+1}\alpha_{j_i}^2,\quad \quad \quad if  \quad card (\{q-q'\})=0 (\Leftrightarrow q=q')\\
\alpha_{q-q'}\alpha_{q'-q},\quad \quad \qquad if \quad card
 (\{q-q'\})=1\\
0, \quad otherwise.
\end{array}\right.
\end{equation}
It is quite simple to notice that the matrices $\mathcal{A}$ and
$\mathcal{B}$ are symmetric. For example,  if $p=1$ ($p=N-2$) the
matrix elements of $\mathcal{A}$, ($\mathcal{B})\in M_N[C]$
assume the following simple form
\begin{equation}\hspace{-2cm}\label{ad}
\mathcal{A}_{ij}= \left\{\begin{array}{c}
\Delta^2+\sum\limits_{l\neq i}^N\alpha_l^2\quad if \quad i=j\\
\alpha_{i}\alpha_{j}\qquad if \quad  i\neq j
\end{array}\right.
,\mathcal{B}_{ij}=\left\{\begin{array}{c}
\Delta^2+\sum\limits_{j=1}^N\alpha_j^2-\alpha_{N+1-i}^2,\quad if
\quad
 i= j,\\
\alpha_{N+1-i}\alpha_{N+1-j},\qquad \qquad if \quad i\neq j.
\end{array}\right.
\end{equation}
Exploiting  the well known solution of a matrix second order
initial-value equation\cite{Libro} like eq. (\ref{A}) yields:
\begin{equation}\label{solution1}
\mathbf{x}(t)=\left[\sum_{k=0}^{+\infty}\frac{(-1)^kt^{2k}}{(2k)!}\mathcal{X}^k\right]\mathbf{x}(0)+\left[\sum_{k=0}^{+\infty}\frac{(-1)^kt^{2k+1}}{(2k+1)!}\mathcal{X}^k\right]\dot{\mathbf{x}}(0)
\end{equation} Moreover, if $\mathcal{X}$ is a non singular
matrix, taking into account that $\mathcal{X}$ admits nonsingular
square roots,
 the solution (\ref{solution1}) may be rewritten in the following closed form
\begin{equation}\label{closed}
\mathbf{x}(t)=\cos\left(\sqrt{\mathcal{X}}\,t\right)\mathbf{x}(0)+\sin\left(\sqrt{\mathcal{X}}\,t\right)(\sqrt{\mathcal{X}})^{-1}\mathbf{\dot{x}}(0).
\end{equation}
 Practically, the possibility of exploiting the solution
(\ref{solution1})/(\ref{closed}) depends of course on our ability
of diagonalizing the  matrices $\mathcal{A}$ and/or $\mathcal{B}$.
The dimensions $C_N^p$ and $C_N^{p+1}$ and the structure of the
two matrices $\mathcal{A}$ and $\mathcal{B}$ respectively make
anyway difficult an analytical treatment. From this point of view,
 solution (\ref{solution1}) might play a formal role only.
However, it's always possible to recourse to the numerical
diagonalization of the matrices $\mathcal{A}$ and/or
$\mathcal{B}$. We may wonder in this case about the efficiency of
this numerical treatment in comparison with other numerical
procedures\cite{Jing}-\cite{Schliemann}. We observe however that
such treatments get effective results when $N$ is confined to
values within $10$, more or less, mainly due to  exponentially
increasing resources required from such computations. Overcoming
these limitations on $N$ becomes thus a mandatory target to
improve the quality of the results achieved from a numerical
approaches. The calculation scheme introduced by us is based on
numerical diagonalization of the matrices $\mathcal{A}$ and
$\mathcal{B}$ and presents the advantage that it enables the
successful treatment of central spin models possessing a large
number of bath spins, at least for small or large polarization,
that is for small or large $p$. Indeed, in these cases the
dimensions of the companion matrices $\mathcal{A}$ and
$\mathcal{B}$ is such that the performance of numerical
simulations is not obstructed by computational obstacles. We
believe that  our diagonalization procedure might be of help to
investigate the behavior of our system as a function of other
initial conditions.
\section{Conclusions}
Over the last few years, a whole variety of methods has been
applied to study decoherence phenomena in the central-spin models.
The appropriate version of the Bethe ansatz   has allowed to
integrate Hamiltonian model, but the results are rather formal and
practically very difficult to handle. In this paper we have
analyzed the exact dynamics of a central spin nonuniformly coupled
through the Heisenberg interaction to a surrounding environment
composed by $N$ spins. Considering arbitrary initial conditions we
have determined  a numerically manipulative general solution from
which information about the full dynamics of our spin models may
be extracted.
\section{Acknowledgments}
One of us (M.A.J.) acknowledges financial support from Fondazione
Don Giuseppe Puglisi "E se ognuno fa qualcosa".

\end{document}